\documentclass[usenatbib]{mn2e}  
\def\lapp{\ifmmode\stackrel{<}{_{\sim}}\else$\stackrel{<}{_{\sim}}$\fi}
\def\gapp{\ifmmode\stackrel{>}{_{\sim}}\else$\stackrel{<}{_{\sim}}$\fi}

\title[Polarization observations of 20 millisecond pulsars]
{Polarization observations of 20 millisecond pulsars}

\author[W. M. Yan et al.]  {W. M. Yan$^{1,2}$\thanks{E-mail:
    yanwm@uao.ac.cn}, R. N. Manchester$^{3}$, W. van Straten$^4$,
  J. E. Reynolds$^3$, G. Hobbs$^3$, 
\newauthor N. Wang$^{1}$, M. Bailes$^4$, N. D. R. Bhat$^4$, S. Burke-Spolaor$^{4,3}$, 
D. J. Champion$^{3,5}$,  
\newauthor W. A. Coles$^6$, A. W. Hotan$^{4,7}$, J. Khoo$^{3}$, S. Oslowski$^{4,3}$,
  J. M. Sarkissian$^3$,
\newauthor J. P. W. Verbiest$^{4,5}$ and D. R. B. Yardley$^{8,3}$ \\
  $^{1}$Urumqi Observatory, NAOC, 40-5 South Beijing Road, Urumqi, Xinjiang, China, 830011\\
  $^{2}$Graduate University of Chinese Academy of Sciences, 19A Yuquan
  Road, Beijing, China, 100049 \\
  $^{3}$CSIRO Astronomy and Space Science, Australia Telescope
  National Facility, PO Box 76, Epping NSW 1710, Australia \\
  $^4$Centre for Astrophysics and Supercomputing, Swinburne University
  of Technology, PO Box 218, Hawthorn VIC 3122, Australia \\
  $^5$Max-Planck-Institut f\"{u}r Radioastronomie, Auf dem H\"{u}gel
  69, 53121 Bonn, Germany \\
  $^6$Electrical and Computer Engineering, University of California at
  San Diego, La Jolla, California, USA \\
  $^7$Curtin Institute of Radio Astronomy, Curtin University, Bentley, WA 6102, Australia \\
  $^8$Sydney Institute for Astronomy, School of Physics, The University 
  of Sydney, NSW 2006, Australia \\
}

\date{\today}

\pagerange{\pageref{firstpage}--\pageref{lastpage}} \pubyear{2010}

\usepackage{psfig}   
\usepackage{threeparttable} 

\DeclareMathAlphabet{\mathbi}{\encodingdefault}{\rmdefault}{\bfdefault}{\itdefault}
\renewcommand\vec[1]{\ensuremath{\mathbi{#1}}}

\begin{document}

\label{firstpage}  

\maketitle

\begin{abstract}
  Polarization profiles are presented for 20 millisecond pulsars that
  are being observed as part of the Parkes Pulsar Timing Array
  project. The observations used the Parkes multibeam receiver with a
  central frequency of 1369 MHz and the Parkes digital filterbank
  pulsar signal-processing system PDFB2. Because of the large total
  observing time, the summed polarization profiles have very high
  signal/noise ratios and show many previously undetected profile
  features. Thirteen of the 20 pulsars show emission over more than
  half of the pulse period. Polarization variations across the
  profiles are complex and the observed position angle variations are
  generally not in accord with the rotating-vector model for pulsar
  polarization. Never-the-less, the polarization properties are
  broadly similar to those of normal (non-millisecond) pulsars,
  suggesting that the basic radio emission mechanism is the same in
  both classes of pulsar. The results support the idea that radio
  emission from millisecond pulsars originates high in the pulsar
  magnetosphere, probably close to the emission regions for
  high-energy X-ray and gamma-ray emission.  Rotation measures were
  obtained for all 20 pulsars, eight of which had no previously
  published measurements.

\end{abstract}


\section{Introduction} \label{sec:intro}

Soon after the discovery of pulsars, it was shown that the radio
emission from pulsars is highly polarized \citep{ls68}. The mean pulse
profile and polarization properties of a pulsar are important for
understanding the pulse emission mechanism, the beaming of pulsar
radiation and the geometry of the system. Mean pulse profiles often
have a double or triple form, leading to a description of the emission
in terms of ``core'' and ``conal'' beams \citep{bac76}. \citet{ran83}
suggested that the core and conal components had different emission
mechanisms whereas \citet{lm88} argued that there was a gradual change
in emission characteristics from the core region to the outer edge of
the emission beam rather than distinct emission processes. To explain
more complex pulse profiles, multiple emission cones were proposed and
discussed by several authors \citep{ran93,kra94,gg03}. An alternative
model suggests that the emission beam of a pulsar is filled with
randomly distributed emission patches \citep{lm88,man95b,hm01}.  The
patchy model may explain more complicated multi-component and
asymmetric pulse profiles.

Polarization properties of radio sources are normally described in
terms of the Stokes parameters: $I$ (total intensity), $Q$ and $U$
(linear polarization) and $V$ (circular polarization). Many pulsars
show a systematic variation of the position angle (PA, $\psi$) of the
linearly polarized emission,
\begin{equation}
\psi = 0.5 \tan^{-1} (U/Q),
\end{equation}
across the pulse profile. In many pulsars the observed PA variations
can be approximately described by the ``Rotating Vector Model" (RVM)
\citep{rc69a}. This model originates from the idea that the radiation
is polarized in the plane of curvature of field lines emanating from a
magnetic pole on the star; not surprisingly, curvature radiation has
this property \citep{kom70}. For a simple dipole field, the observed
PA variation is then determined by the projected direction of the
magnetic axis as the star rotates. The rapid swings often observed
near the profile mid-point imply that magnetic axis is nearly aligned
with the observer's line of sight at that profile phase. Circular
polarization (Stokes $V$) is usually relatively weaker than linear
polarization. It is most often associated with the central or core
component of the profile, often with a sense reversal near the profile
mid-point \citep{ran83}. Observed PA swings are not always smooth and
continuous. Discontinuities of approximately 90\degr~are often
observed \citep[e.g.,][]{mth75,br80,scr+84,hdvl09} and these are
interpreted as resulting from overlapping emission from orthogonally
polarized emission modes \citep[e.g.,][]{ms98}. Such orthogonal modes
may be generated as the wave propagates through the pulsar
magnetosphere \citep[e.g.,][]{pet01}.

In many respects, millisecond pulsars (MSPs) have very different
properties to those of ``normal'' pulsars. Besides their obviously
shorter periods, they have much smaller period derivatives and hence
larger characteristic ages and weaker implied dipole magnetic
fields. The pulse profile for millisecond pulsars generally has a much
larger duty cycle, i.e., the emission covers a much larger fraction of
the pulse period, and a more complicated pulse profile compared to
normal pulsars \citep[see, e.g.,][]{kxl+98,nms+97}. This might be
expected within the context of the magnetic-pole model since the
pulsar magnetosphere, bounded by the light cylinder of radius $R_{LC}
= cP/2\pi$ is so much smaller.  Consequently, the polar caps defined
by the open field lines (those that penetrate the light cylinder) are
larger and the opening angle of these open field lines at a given
radial distance $r$ and fractional polar-cap radius $s$ is also larger:
\begin{equation}\label{eq:beam}
\theta_b \approx \frac{3}{2}s\left(\frac{r}{R_{LC}}\right)^{1/2}
\end{equation}
where $\theta_b$ is the field-line tangent angle with respect to the magnetic
axis (see e.g., \cite{dwd10}). Despite these
differences, the polarization characteristics of MSPs are remarkably
similar to those of normal pulsars -- they have similar high degrees
of linear polarization, orthogonal-mode PA jumps are observed and
circular polarization with sense changes near the profile centre is
seen in some MSPs \citep[see,
e.g.,][]{ts90,nms+97,stc99,ovhb04,mh04}. Although the emission
characteristics are similar, indicating that the emission mechanism is
basically the same for normal pulsars and MSPs, the PA variations in
MSPs are often more complex and do not fit the RVM well. 

Faraday rotation occuring in the interstellar medium is 
relatively easy to observe in pulsars because of the typically high
degree of linear polarization. Faraday rotation is quantified by the
rotation measure (RM), defined by
\begin{equation}
\psi = {\rm RM}\; \lambda^2,
\end{equation} 
where $\lambda = c/\nu$ is the radio wavelength
corresponding to radio frequency $\nu$. Faraday rotation in the
interstellar medium is given by:
\begin{equation}
{\rm RM} = 0.810 \int_0^D n_e \vec{B \cdot dl}
\end{equation}
where $\vec{B}$ is the vector magnetic field in microgauss and the
integral is over the path to the pulsar, and so
observations of Faraday rotation can be used to investigate the
Galactic magnetic field \citep[see,
e.g.,][]{man74,hml+06,njkk08}. Pulsars have the unique advantage that
the dispersion measure
\begin{equation}
{\rm DM} = \int_0^D n_e\;dl,
\end{equation}
is also known. We can therefore obtain a measure of the mean
line-of-sight component of the interstellar magnetic field weighted by
the local electron density along the path:
\begin{equation}\label{eq:bpar}
\langle B_{||}\rangle = 1.232\; \frac{\rm RM}{\rm DM}\;\mu{\rm G}
\end{equation}
where RM and DM are in their usual units. 

In this paper we report on the mean pulse profiles and polarization
properties of the 20 MSPs which are being observed as part of the
Parkes Pulsar Timing Array (PPTA) project
\citep{man08,hbb+09}. Details of the observing system and the
observations are given in \S\ref{sec:obs} and polarization profiles
for the 20 pulsars are presented in \S\ref{sec:poln}. RM results are
presented in \S\ref{sec:rm} and the implications of the results are
discussed in \S\ref{sec:discn}.

\section{Observations and Analysis} \label{sec:obs} 

As part of the PPTA project, frequent observations of 20 MSPs are made
using the Parkes 64-m radio telescope. Regular observations at
approximately three-weekly intervals commenced in mid-2004. Since that
time, back-end signal processing systems have been upgraded several
times. The main objective of these observations is to detect
gravitational waves through high-precision pulsar timing. However,
there are many secondary objectives, including studies of the pulse
emission properties for pulsars in the observed sample. All PPTA data
are recorded in full-polarization mode, not only enabling full profile
calibration, but also making possible detailed studies of the pulse
emission properties. This paper reports on polarization results from
observations made with the centre beam of the Parkes 20cm Multibeam
receiver \citep{swb+96} with the second-generation Parkes digital
filterbank system PDFB2 between 2007 June and 2009 November. The fact
that many observations are available for all 20 pulsars means that
pulse profiles with very high signal/noise ratios can be formed,
leading to new information about the profile shapes and their
polarization.

Feeds in the Parkes 20 cm Multibeam receiver have orthogonal linearly
polarized probes with a calibration probe at 45\degr~to the two signal
probes through which a linearly polarized broad-band and pulsed
calibration signal can be injected. The system
equivalent flux density on cold sky is approximately 30 Jy. In common
with the other Parkes digital filterbank systems, PDFB2 digitizes
band-limited signals from each of the two orthogonal probes at the
Nyquist rate with 8-bit sampling. It uses field-programmable gate array 
(FPGA) processors to implement
a polyphase filter to provide frequency resolution and a synchronous
averager to form mean pulse profiles. For the observations reported
here the total bandwidth was 256 MHz centred at 1369 MHz with 1024
channels across the band for most of the 20 pulsars.  For all except
PSRs J1857+0943, J1939+2134 and J2124$-$3358, each observation was of 64
minutes duration; for the exceptions, the observation times were 32
minutes. All data were recorded using the PSRFITS data format
\citep{hvm04} with 1-minute sub-integrations and the full spectral
resolution. To provide a calibration of the gain and phase of the
receiver system, the pulsed calibration signal was recorded for 2
minutes prior to each pulsar observation. Signal amplitudes were
placed on a flux density scale using observations of Hydra A, assumed
to have a flux density of 43.1 Jy at 1400 MHz and a spectral index of
$-$0.91. The 20cm Multibeam feeds have limited isolation between the
orthogonal polarizations. This cross-coupling was measured using
series of observations of PSR J0437$-$4715 covering a wide range of
hour angles, allowing correction for the effects of this coupling
\citep{van04c}.

Off-line data processing made use of the PSRCHIVE pulsar data analysis
system \citep{hvm04}. First, band edges (5\% on each side) and
spurious data resulting from narrow-band and impulsive radio-frequency
interference were effectively excised from the raw data
files. Typically 1 -- 2\% of the spectrum and the time-domain data
were excised. The data were then averaged in time to give eight
sub-integrations for each observation. Next, the data for each
observation were corrected for variations in instrumental gain and
phase across the spectrum and for the effects of cross-coupling in the
feed. Finally, baselines for each of the Stokes-parameter profiles
were set to zero mean and the data placed on a flux density scale. The
baseline region was computed from the Stokes $I$ profile and applied to
all four Stokes-parameter profiles. Following this calibration,
Stokes parameters were in accordance with the astronomical conventions
described by \citet{vmjr10}. Specifically, PAs are absolute and are
measured from celestial North toward East, i.e., counterclockwise on
the sky, and Stokes $V$ is defined as $I_{LH}-I_{RH}$, using the IEEE
definition for sense of circular polarization.

In order to form final mean polarization profiles, it is necessary to
sum across the frequency channels taking into account the Faraday
rotation across the band. Sufficiently precise RM measurements were
not previously available for most of the PPTA sample. We therefore
determined RM values for each pulsar from the PDFB2 data sets.  For
each individual observation, the upper and lower halves of the total
band were separately summed using the nominal RM value. An improved value
of the RM was then calculated by taking a weighted mean of position
angle differences for bins where the uncertainty in position angle was
less than 10\degr. This process was iterated until convergence was
obtained. The contribution of the Earth's ionosphere to the total RM
was estimated and then subtracted so that the measured RM just
represents the interstellar contribution. Ionospheric RMs were
computed using a prediction program {\sc farrot}, developed at the
Dominion Radio Astrophysical Observatory, Penticton, which uses the
observed Solar 10.7-cm radio flux as a basic input. As the
observations were made near a minimum of Solar activity, ionospheric
RMs are relatively small, typically between $-0.5$ and
$-1.0$~rad~m$^{-2}$. The final interstellar RM for each pulsar was
determined by taking a weighted mean of the individual interstellar RM
measurements. In addition, the observed Stokes parameters for each
observation were adjusted (effectively a PA shift) using the
ionospheric RM so that they represent the polarization state at the
top of the ionosphere. 

It is also necessary to have a precise timing model for each pulsar
before adding the data in time to form a final mean profile. Pulse
times of arrival were obtained for each observation using an
analytic template based on an existing high signal/noise (S/N) ratio
pulse profile. The TEMPO2 pulsar timing package \citep{hem06} was then
used to fit pulsar spin and astrometric and binary parameters as
necessary to give ``white'' timing residuals for the PDFB2 data set
for each pulsar. Finally, the separate observations were summed using
this timing model to determine relative phases to form the final
Stokes parameter profiles. To give the best possible S/N ratio in the
final profile, the individual observation profiles were weighted by
their (S/N)$^2$ when forming the sum profile. The polarization
parameters $L = (Q^2 + U^2)^{1/2}$, $|V|$ and PA were then computed
from the Stokes parameters. The noise bias in $L$ was corrected using
the relations given by \citet{lk05} and the similar bias in $|V|$ was
corrected using
\begin{equation}
|V|_{\rm corr} = (|V|^4 - \sigma_I^4)^{1/4}
\end{equation}
when $|V| > \sigma_I$ or zero otherwise, where $\sigma_I$ is the rms
profile baseline noise.

Table~\ref{tb:obs} gives the basic pulsar and observational parameters
for the 20 PPTA pulsars which were observed. After the pulsar name,
pulse period and DM, we give the number of frequency channels across
the 256-MHz band and the number of bins across the pulse period. The
dispersion smearing across each frequency channel, given by
\begin{equation}
  \Delta t_{\mathrm{DM}} \approx 8.30 \times 10^6 \;\mathrm{DM}\;\Delta f\;f^{-3}\;\;\mathrm{ms},
\end{equation}
where $\Delta f$ is the channel width in MHz, $f$ is the band centre
frequency in MHz, and the DM is in units of cm$^{-3}$pc is given in
units of profile bins, the number of observations summed, and 
the total observation time are given in the remaining three channels.

\begin{table*}
\begin{center}
\centering
\caption{Observational parameters for the 20 PPTA MSPs}
\label{tb:obs}
\begin{threeparttable}
\begin{tabular}{cccccccc}
\hline
{PSR} & $P$  & DM  & Nr of  &Nr of & DM Smear & Nr of &Integ. Time  \\
	    &(ms)  &(cm$^{-3}$ pc)    & Channels  & Bins &(bins) & Obs. &(h) \\
\hline                                                                             
J0437$-$4715     &5.757 &2.64      & 1024    &1024 &0.4  & 92 &96.3\\
J0613$-$0200     &3.062  &38.78    &  1024  & 512  &5.2 &41  &43.3\\
J0711$-$6830     &5.491  &18.41    &  1024 & 512 &1.4 &28  &29.1\\
J1022+1001       &16.453 &10.25   &1024      & 2048 &1.0 &34 &34.3\\
J1024$-$0719     &5.162   &6.49       &1024     & 1024 &1.0 &29  &30.0\\\\
J1045$-$4509     &7.474   &58.17    &2048    & 512 &1.6 &33 &34.1\\
J1600$-$3053     &3.598  &52.33   &1024   & 512  &6.0 &28 &29.7\\
J1603$-$7202    &14.842  &38.05    & 1024  &1024 &2.1 &23 &23.7\\
J1643$-$1224    &4.622   &62.41    &1024  &  512 &2.8 &34 &36.1\\
J1713+0747     &4.570    &15.99   &1024    &1024 &2.9 &40  &40.7\\\\
J1730$-$2304     &8.123  &9.62    &1024    & 1024 &1.0 &23 &23.7\\
J1732$-$5049   &5.313     & 56.82   &  2048    & 512 &2.2 &24  &24.7\\
J1744$-$1134   &4.075    &3.14    &512     & 1024 &1.3  &33  &34.4\\
J1824$-$2452    &3.054   &120.50    &2048    & 256 &4.1  &27 &28.0\\
J1857+0943       &5.362  &13.30   &1024   & 1024 &2.1  &26 &13.4\\\\
J1909$-$3744    & 2.947  &10.39    &1024   & 512 &1.5  &67 &69.6\\
J1939+2134       & 1.558 &71.04   & 1024  &256  &9.4  &26 &13.8\\
J2124$-$3358     & 4.931 &4.60     & 1024    &1024 &0.8  &25  &13.1\\
J2129$-$5721     & 3.726 &31.85    & 1024    & 512 &3.5 &25  &25.5\\
J2145$-$0750     &16.052 &9.00    & 1024   & 2048 &0.9  &29  &29.6 \\
\hline
\end{tabular}
\end{threeparttable}
\end{center}
\end{table*}

\section{Polarization profiles}\label{sec:poln}
Polarization profiles at 1369 MHz for the 20 PPTA MSPs are presented
and discussed in this section. Table~\ref{tb:poln} gives a summary of
the results. Columns 2 and 3 give the mean flux density $S = \langle I
\rangle$ averaged over all observations and its uncertainty. The next
column gives the rms fluctuation in individual-observation flux
densities. For low DM pulsars (DM $\lapp 30$~cm$^{-3}$~pc) the rms
fluctuation is comparable to or even greater than the mean value. This
entirely results from intensity modulations due to diffractive
interstellar scintillation. Pulse widths at 50\% of the peak flux
density (W50) and 10\% of the peak (W10) are given next, both in
degrees of longitude, where $360\degr$ is equivalent to the pulse
period or 1.0 in pulse phase, and in milliseconds. The next column
gives the overall pulse width, measured from the first point to the
last point where the pulse intensity significantly exceeds the
baseline noise (more than three times the baseline rms noise in
several adjacent bins) . The next column gives the minimum number of
identifiable pulse components in the pulse profile. A component is
identified by a peak or, for overlapping components a significant
inflection, in the pulse profile. The next three columns give the
fractional linear polarization $\langle L \rangle/S$, the fractional
net circular polarization $\langle V \rangle /S$ and the fractional
absolute circular polarization $\langle |V| \rangle/S$ averaged over
all observations, where the $\langle \rangle$ means are taken across
the pulse profile.

\begin{table*}
\begin{center}
\centering
\caption{Flux density and polarization parameters for PPTA pulsars}
\label{tb:poln}
\begin{tabular}{cccccccccccccc}
\hline
{PSR}  & $S$ & $\sigma_S$& $S_{\rm RMS}$ &\multicolumn{2}{c}{W50}&  &\multicolumn{2}{c}{W10} & Overall Width & Nr of & 
$\left.\left\langle{L}\right\rangle\middle/S\right.$  & $\left.\left\langle{V}\right\rangle\middle/S\right.$ &$\left.\left\langle\left|{V}\right|\right\rangle\middle/S\right.$   \\
\cline{5-6}
\cline{8-9}
	   &(mJy) &(mJy)  &(mJy) &(deg) &(ms)  &  &(deg) &(ms) & (deg) & Comp. &(\%)  &(\%) & (\%)  \\

\hline
J0437$-$4715  &149.3 &4.0  &38.0 &9&0.14 &  &64 &1.02   & 306 & 12 &24    &$-$5   & 11   \\
J0613$-$0200   &2.3  &0.1  &0.6  &55 &0.92  & &108 &8.18  &158 &  7 &15    &2      & 4    \\
J0711$-$6830   &1.4  &0.3  &2.0  &124&1.90 & &169 &2.57  &277  & 10 &4     &$-$3  & 4   \\
J1022+1001     &1.5  &0.3  &2.0  &21 &0.98   & &43   &1.96 &68  &  6 &51    & $-$11 & 14  \\
J1024$-$0719   &1.5  &0.3  &1.7  &35  &0.50 & &106 &1.51  &158  &  9 &55    & 2    & 7   \\ \\
J1045$-$4509   &2.2  &0.1  &0.5  &37  &0.76 & &70   & 1.45&248  &  9 &20    & 13   & 14   \\
J1600$-$3053   &2.4  &0.1  &0.4  & 9 &0.09 &   &41   &0.42 &68  &  3 &27    & 2   & 3 \\
J1603$-$7202   &4.2  &0.7  &3.4  &29  &1.21& &41   &1.71 &238 &  7 &17    & 27 & 28   \\
J1643$-$1224   &5.0  &0.7  &4.2  &25  &0.32 & &72  &0.93 &241  &  5 &11    &$-$1 & 9      \\
J1713+0747     &7.4  &1.1  &7.4  &9   &0.11   &  &30  &0.39 &104  &  5 &28    & 0   & 3   \\ \\
J1730$-$2304   &3.9  &0.3  &1.5  &43&0.98  & &77&1.74  &232  &  9 &27    &$-$20   &20  \\
J1732$-$5049   &1.3  &0.1  &0.5  &20  &0.29& &119  &1.76 &155 &  4 &7    & 0   & 2 \\
J1744$-$1134   &3.3  &0.6  &3.5  &12  &0.14&  &22   &0.25 &209 &  8 &87    &  1   & 3 \\
J1824$-$2452   &1.6  &0.1  &0.6  &115 &0.98 & &189 &1.60  & 281 &  6 &58    &  1   &  6  \\
J1857+0943     &5.9  &0.8  &4.1  &35  &0.52&   &203  &3.03 &238 &  6 &4      & 2 & 2 \\ \\
J1909$-$3744   &2.6  &0.2  &2.0  &5   &0.04&   &11   &0.09 &187 &  3 &47    & 12   &13    \\
J1939+2134     &13.8 &0.9  &4.6  &15 &0.06&   &198 &0.86  & 302 &  6 &27    & 1  &  2 \\
J2124$-$3358   &2.4  &0.4  &2.1  &38 &0.53&   &275 &3.76 &331  & 12 &18    & 0    & 3   \\
J2129$-$5721   &1.6  &0.1  &0.7  &26 &0.27&   &60   &0.62 &144 &  5 &41    & $-$23   & 26  \\
J2145$-$0750   &9.3  &1.5  &7.9  &8  &0.34&  &94 &4.17 &187 &  7 &10    & 6   &  6  \\
\hline
\end{tabular}
\end{center}
\end{table*}

In the following subsections, we present polarization profiles and discuss each pulsar in turn.

\subsection{PSR J0437$-$4715}\label{J0437}

\begin{figure}
\centerline
{
 \psfig{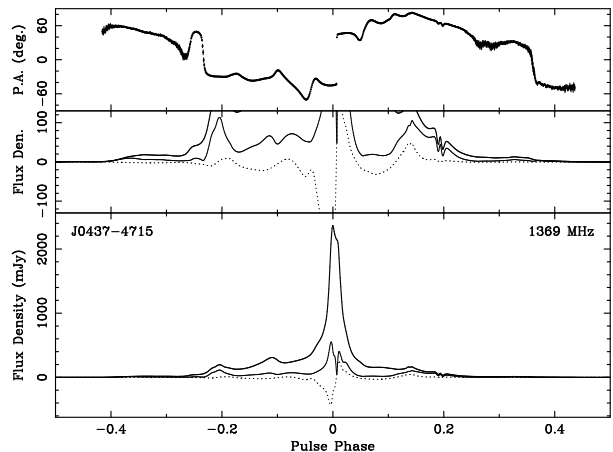}
}
\caption{Polarization profiles for PSR J0437$-$4715 at 1369 MHz. The
  lower part gives the pulse profile for total intensity $I$ (thick
  solid line), linearly polarized intensity $L$ (thin solid line), and
  circularly polarized intensity $V$ (dotted line). The middle part is
  an expanded plot showing low-level details of the polarization
  profiles and the upper part gives the position angle of the linearly
  polarized emission.}
\label{Fig:0437}
\end{figure}

Mean pulse profile and polarization parameters for this very strong
southern MSP (mean flux density about 145 mJy, Table~\ref{tb:poln}) at
1369 MHz are given in Fig.~\ref{Fig:0437}.  Our results are in good
agreement with and extend previously published results
\citep{jlh+93,mj95,nms+97}. As is well known, the total intensity and
polarization variations across the pulse profile are complex. There
are multiple overlapping components and the present results show that
the pulse profile extends over at least 85\% of the pulse period. The
notches in Stokes $I$ and $L$ discussed by \citet{nms+97} and
\citet{drr07} are clearly visible in the central expanded plot. The
observed PA variations cannot be described by the RVM, even if
orthogonal mode jumps are taken into account. There is a clear
orthogonal mode transition in both linear and circular polarization
very close to the main profile peak and a non-orthogonal PA transition
around pulse phase $-0.23$ \citep{mj95,nms+97}. We also observe a
probable non-orthogonal transition near pulse phase 0.35.

\subsection{PSR J0613$-$0200}

\begin{figure}
\centerline
{
 \psfig{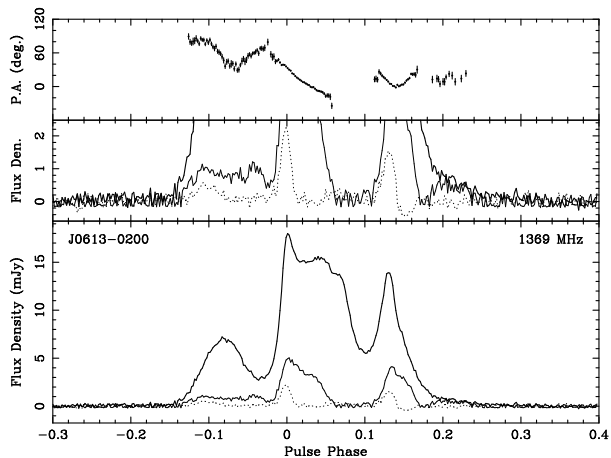}
}
\caption{Polarization profiles for PSR J0613$-$0200 at 1369 MHz. See
  Fig.~\ref{Fig:0437} for further details.}
\label{Fig:0613}
\end{figure}

Fig.~\ref{Fig:0613} shows the polarization profiles for PSR
J0613$-$0200. These are consistent with and extend the 1341 MHz
results presented by \citet{ovhb04}.  \citet{stc99} observed very
different profiles at 410 MHz and 610 MHz,\footnote{Note that
  \citet{stc99} uses a reversed PA sign compared to the IAU convention
  used here.} indicating that there is significant profile morphology
evolution between these lower frequencies and the 1369 MHz profile
presented here. In particular, the trailing peak is much stronger at
low frequencies relative to the rest of the profile. Our results have
higher S/N ratio than those previously published, and show that the
pulse profile consists of multiple overlapping components; a weak
trailing component was not obvious in previous results. Significant
linear polarization is seen across essentially the whole profile,
showing that the PA variations are more complex than was evident from
earlier work, with several regions of increasing and decreasing PA
across the profile. There is evidently a weak pulse component at the
leading edge of the trailing component (at phase 0.12) which has a
distinctly different PA, discontinuous with the rest of the trailing
component. Two clear peaks in circular polarization are seen, under
the leading edges of the central broad pulse peak and the trailing
peak. There is a sense reversal in the circular polarization within
the trailing component.

\subsection{PSR J0711$-$6830}

\begin{figure}
\centerline
{
 \psfig{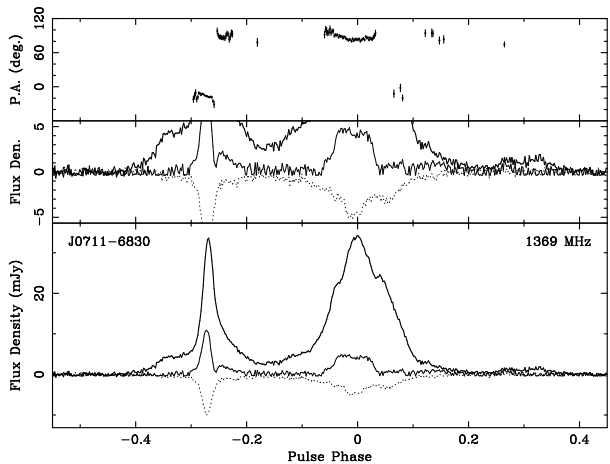}
}
\caption{Polarization profiles for PSR J0711$-$6830 at 1369 MHz. See
  Fig.~\ref{Fig:0437} for further details.}
\label{Fig:0711}
\end{figure}

The mean pulse profile for this pulsar is basically a wide double with
a connecting bridge of emission \citep{ovhb04,mh04}. However, the
results of \citet{ovhb04} showed that there were several overlapping
components in the main peaks of the profile and furthermore that there
was evidence for weak emission following the second main component,
with significant emission over 75\% of the pulse period. These
features are confirmed by the observations shown in
Fig.~\ref{Fig:0711}. As well as the component preceding the first
peak, there is clear emission following the second peak, with a low
double-peaked feature close to phase 0.3 in Fig.~\ref{Fig:0711}. The
polarization properties from our observations are very similar to
those of \citet{ovhb04} with both linear and circular peaks centred
under the main components. However, we find evidence for an orthogonal
mode transition just after the peak of the leading main component.

\subsection{PSR J1022+1001}\label{J1022}

\begin{figure}
\centerline
{
 \psfig{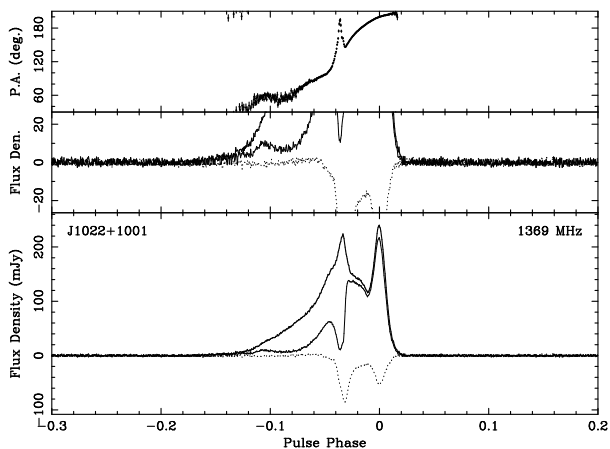}
}
\caption{Polarization profiles for PSR J1022+1001 at 1369 MHz. See
  Fig.~\ref{Fig:0437} for further details.}
\label{Fig:1022}
\end{figure}

Mean pulse profile and polarization parameters for this pulsar at 1369
MHz are given in Fig.~\ref{Fig:1022}. The polarization profiles shown
here are generally in good agreement with those presented by
\citet{kxc+99}, \citet{stc99} and \citet{ovhb04} except that the
fractional linear polarization in the trailing component in our
observations is a little higher. Given that the overall PA variation
closely fits the RVM, it seems probable that the PA glitch near the
centre of the profile results from a separate emission
component. Similar central PA discontinuities are seen in several
pulsars, e.g. PSR B1913+16 \citep{bcw91} and PSR J1141$-$6545
\citep{mks+10}. Alternatively, it may result from a pair of closely
spaced and smeared orthogonal-mode transitions.

\subsection{PSR J1024$-$0719}

\begin{figure}
\centerline
{
 \psfig{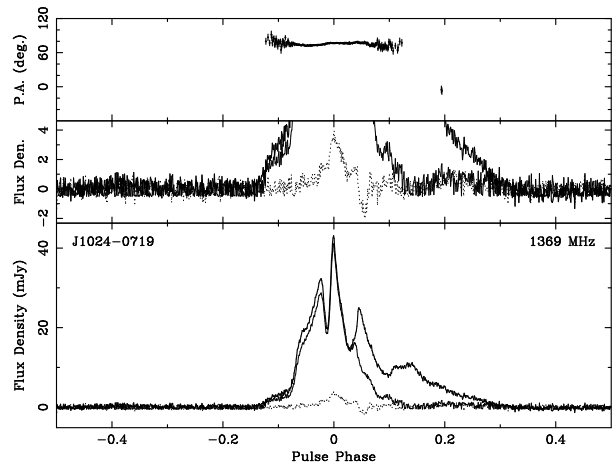}
}
\caption{Polarization profiles for PSR J1024$-$0719 at 1369 MHz. See
  Fig.~\ref{Fig:0437} for further details.}
\label{Fig:1024}
\end{figure}

Our results, presented in Fig.~\ref{Fig:1024}, have a higher S/N ratio
compared to earlier work \citep[e.g.,][]{ovhb04} and show that the
pulse profile has multiple overlapping features. The leading part of
the profile has extremely high fractional linear polarization whereas
the trailing part of the profile is essentially unpolarized. Circular
polarization is weak except near the centre of the profile where there
is a sense reversal. There is sub-structure in the circular
polarization with at least three overlapping components in the
positive $V$ part. In agreement with previous work we find that there
is little or no variation in PA across the profile. 

\subsection{PSR J1045$-$4509}

\begin{figure}
\centerline
{
 \psfig{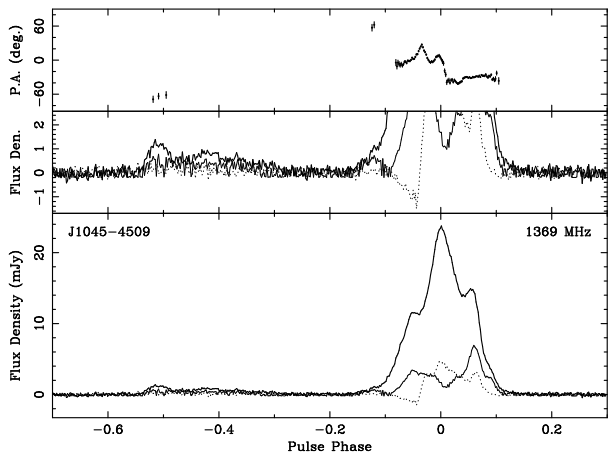}
}
\caption{Polarization profiles for PSR J1045$-$4509 at 1369 MHz. See
  Fig.~\ref{Fig:0437} for further details.}
\label{Fig:1045}
\end{figure}

Relatively low-quality polarization profiles have been presented for
this pulsar by \citet{ovhb04} and \citet{mh04}. The data shown in
Fig.~\ref{Fig:1045} are have higher S/N ratio and show a number of
previously unrecognised profile features. Most notable is the
leading emission around phase $-0.5$ which is joined to the main pulse
by a low-level bridge of emission. This low-level emission extends the
overall pulse width to more than 60\% of the pulse period. The
polarization properties shown in Fig.~\ref{Fig:1045} are consistent
with those given in the earlier papers but show more detail. In
particular, the PA variation is more complex with non-monotonic
variations and a probable orthogonal transition near the leading edge
of the main pulse. There is a circular sense reversal between the
first and second strong components of the main pulse.

\subsection{PSR J1600$-$3053}

\begin{figure}
\centerline
{
 \psfig{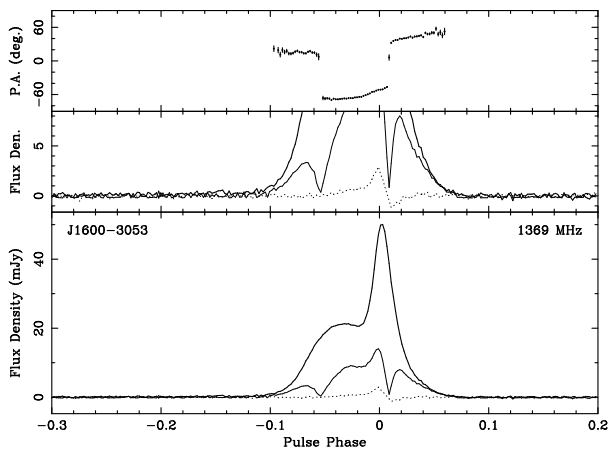}
}
\caption{Polarization profiles for PSR J1600$-$3053 at 1369 MHz. See
  Fig.~\ref{Fig:0437} for further details.}
 \label{Fig:1600}
\end{figure}

Mean pulse profile and polarization parameters for this pulsar at 1369
MHz are shown in Fig.~\ref{Fig:1600}.  Our observations are in
complete agreement with those of \citet{ovhb04}. Two orthogonal PA
transitions are seen. The second of these slightly trails the
strongest profile peak and is coincident with a sense reversal in
$V$. This situation is virtually identical to that seen in PSR
J0437$-$4715, with a fully orthogonal polarization transition very
close to the main pulse peak.

\subsection{PSR J1603$-$7202}

\begin{figure}
\centerline
{
 \psfig{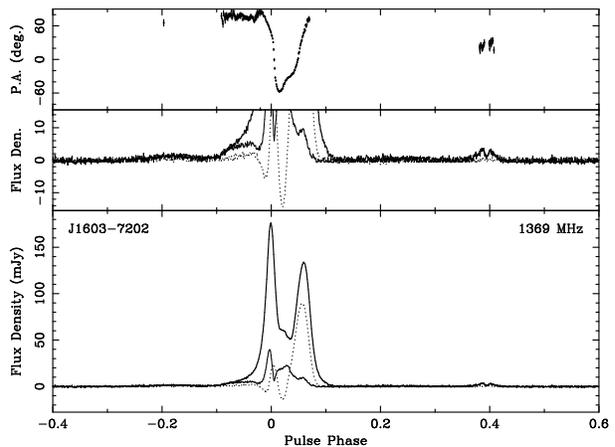}
}
\caption{Polarization profiles for PSR J1603$-$7202 at 1369 MHz. See
  Fig.~\ref{Fig:0437} for further details.}
 \label{Fig:1603}
\end{figure}

Mean pulse profile and polarization parameters presented in
Fig.~\ref{Fig:1603} show that the profile for this pulsar has two main
components, with the trailing one having very high circular
polarization \citep[cf.,][]{ovhb04,mh04}. Table~\ref{tb:poln} shows
that this pulsar has the highest fractional mean $V$ and $|V|$ of the
PPTA pulsars. The expanded central plot in Fig.~\ref{Fig:1603} shows
that there is a broad low-level feature preceding the main pulse and a
double-peaked pulse trailing the main pulse by about 0.4 in
phase. With these additional pulse components, the pulse profile
extends over about 70\% of the period. Variations in polarization
across the pulse are complex with three sense reversals in $V$ and an
unusual PA variation which bears no relation to that expected for
RVM. 

\subsection{PSR J1643$-$1224}

\begin{figure}
\centerline
{
 \psfig{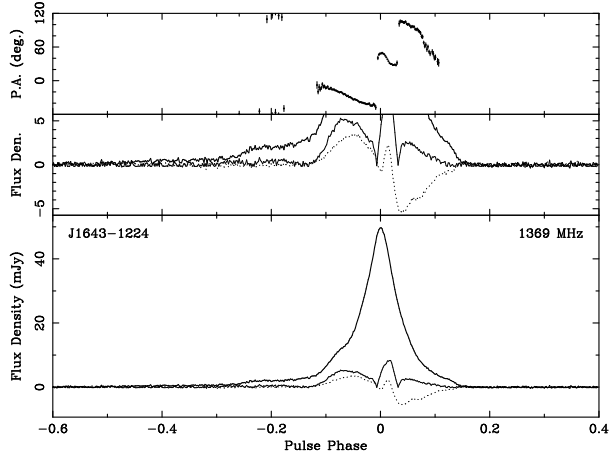}
}
\caption{Polarization profiles for PSR J1643$-$1224 at 1369 MHz. See
  Fig.~\ref{Fig:0437} for further details.}
 \label{Fig:1643}
\end{figure}

The pulse polarization profiles given in Fig.~\ref{Fig:1643} have far
more detail than previously published results
\citep{stc99,mh04,ovhb04}. They show that the pulse profile has a long
leading ramp (phase $-0.6$ to $-0.3$ in Fig.~\ref{Fig:1643}) followed
by a broad component leading into the main pulse which has at least
three overlapping components. There are two orthogonal polarization
transitions near the pulse peak and these are associated with sense
reversals in $V$ and dips in $L$. When account is taken of the two
discontinuities, there is a decreasing PA through the stronger part of
the profile, but with a significantly steeper PA swing in the trailing
part. The PA of the broad feature preceding the main pulse is not very well
determined, but it appears discontinuous with the rest of the PA
variations.

\subsection{PSR J1713+0747}

\begin{figure}
\centerline
{
 \psfig{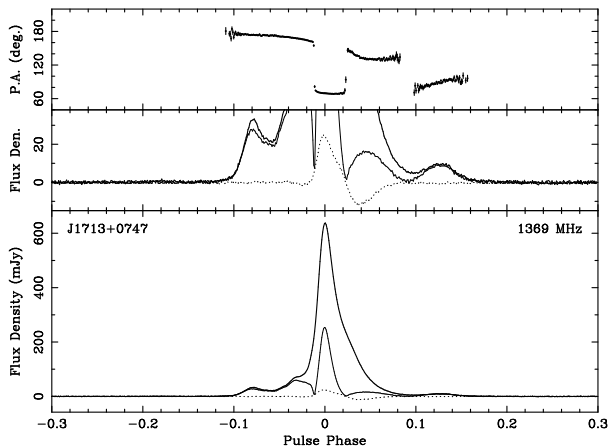}
}
\caption{Polarization profiles for PSR J1713+0747 at 1369 MHz. See
  Fig.~\ref{Fig:0437} for further details.}
 \label{Fig:1713}
\end{figure}

Fig.~\ref{Fig:1713} shows that this pulsar has a multi-component mean
pulse profile extending over a little more than 0.25 in phase. Our
results are consistent with, but improve on, those of
\citet{ovhb04}. Both the leading and trailing parts of the pulse
profile have very high, almost complete, linear polarization. Similar
to PSR J1643$-$1224, there are two orthogonal PA transitions near the
pulse peak. The second one of these is clearly associated with a sense
change in $V$ and the first may be also. A third PA transition is
observed preceding the trailing pulse component. However, this is
clearly not orthogonal, with $\Delta\psi \sim 70\degr$. 

\subsection{PSR J1730$-$2304}

\begin{figure}
\centerline
{
 \psfig{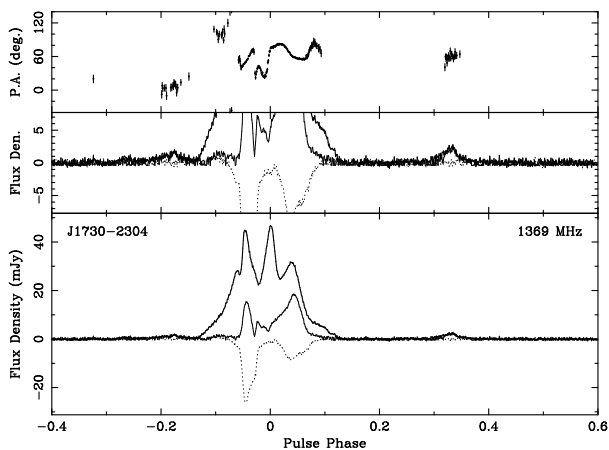}
}
\caption{Polarization profiles for PSR J1730$-$2304 at 1369 MHz. See
  Fig.~\ref{Fig:0437} for further details.}
 \label{Fig:1730}
\end{figure}

Mean pulse profile and polarization parameters for this pulsar at 1369
MHz are given in Fig.~\ref{Fig:1730}. These show that the pulse
profile is very complex, with multiple components in the main part,
two weak leading components (around phases $-0.18$ and $-0.26$) and a
slightly stronger trailing component (around phase 0.32). The weak
trailing component was also observed by \citet{kxl+98} but the leading
components have not been previously identified. This pulsar is unusual
in that it has relatively strong right-circular polarization through
much of the main pulse \citep[cf.,][]{ovhb04}. The observed PA
variation is very complex with a possible orthogonal transition in the
leading part of the main pulse. The rest of the profile shows complex
PA variations with zones of both increasing and decreasing PA. Both
the second precursor component and the postcursor are highly linearly
polarized, probably with increasing PA across them. The leading
precursor is too weak to be sure about its polarization.

\subsection{PSR J1732$-$5049}

\begin{figure}
\centerline
{
 \psfig{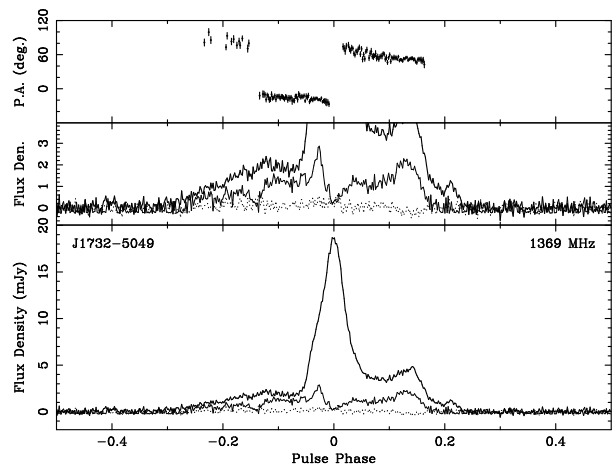}
}
\caption{Polarization profiles for PSR J1732$-$5049 at 1369 MHz. See
  Fig.~\ref{Fig:0437} for further details.}
 \label{Fig:1732}
\end{figure}

PSR J1732$-$5049 is a relatively weak pulsar. Results presented in
Fig.~\ref{Fig:1732} are similar to those given by \citet{ovhb04}
except that the fractional linear polarization is a little higher. No
significant circular polarization is observed across the profile. The
position angle swing is relatively flat with two PA jumps, both close
to orthogonal.

\subsection{PSR J1744$-$1134}

\begin{figure}
\centerline
{
 \psfig{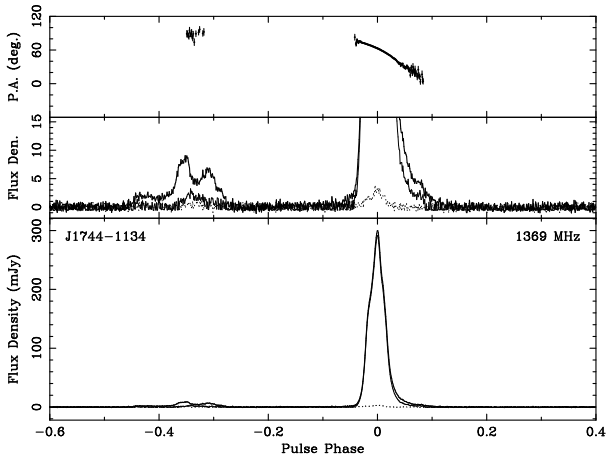}
}
\caption{Polarization profiles for PSR J1744$-$1134 at 1369 MHz. See
  Fig.~\ref{Fig:0437} for further details.}
 \label{Fig:1744}
\end{figure}

As shown in Fig.~\ref{Fig:1744}, the mean pulse profile of this pulsar has
a sharp main pulse and a precursor preceding the main pulse by $\sim
0.35$ in phase. The precursor was first detected by \citet{kxl+98};
these authors also found evidence for a postcursor following the main
pulse by about 0.2 in phase and amplitude about half of that of the
precursor. We do not see this postcursor
component. Fig.~\ref{Fig:1744} clearly shows that the precursor has
multiple components. This pulsar is remarkable in that the main pulse
is almost 100\% linearly polarized, especially on the leading
edge. However weak left-circular emission is observed in both the
precursor and the main pulse. There is a smooth decrease of PA through
the main pulse. This may be continuous with the PA of the precursor,
indicating an outer line of sight, i.e., that the observer's line of
sight is equatorward of the magnetic axis. 

\subsection{PSR J1824$-$2452 (PSR B1821$-$24)}

\begin{figure}
\centerline
{
 \psfig{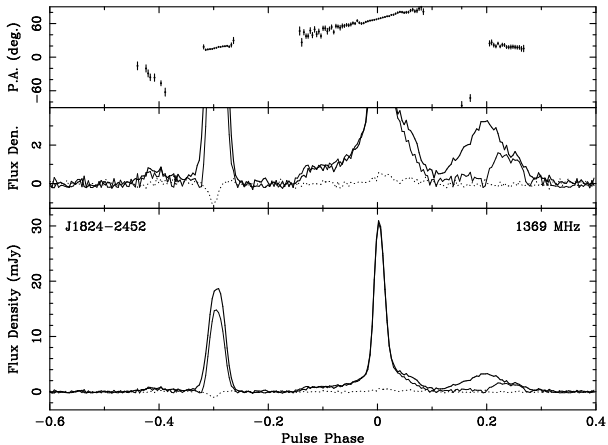}
}
\caption{Polarization profiles for PSR J1824$-$2452 at 1369 MHz. See
  Fig.~\ref{Fig:0437} for further details.}
 \label{Fig:1824}
\end{figure}

As Fig.~\ref{Fig:1824} shows, this pulsar, which is located in the
globular cluster M28 \citep{lbm+87}, has a complex profile with
multiple components covering more than 80\% of the pulse
period. Comparison with the results of \citet{stc99} shows that the
leading strong component in Fig.~\ref{Fig:1824} has a very steep
spectrum as it is by far the strongest component at 610 MHz. Our
results are generally in agreement with those of \citet{ovhb04} but
show more detail, including a weak component around phase $-0.4$. Most
of the strongest component in Fig.~\ref{Fig:1824} is 100\% linearly
polarized. In fact, there is slight overpolarization of the ramp
preceding this component. This may result from the difficulty in
accurately estimating baselines for this very wide profile or from the
existence of polarized unpulsed emission from the pulsar. The
first strong component is also very highly polarized. The PA variation
across these components appears continuous, but the PA of the extreme
leading and trailing components is not continuous with the central
part, having the opposite slope.

\subsection{PSR J1857+0943 (PSR B1855+09)}

\begin{figure}
\centerline
{
 \psfig{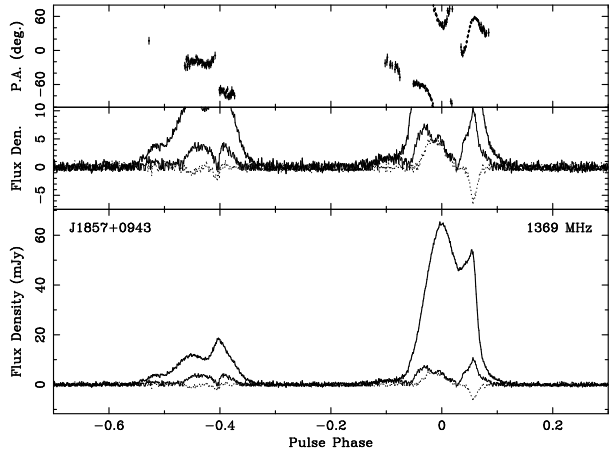}
}
\caption{Polarization profiles for PSR J1857+0943 at 1369 MHz. See
  Fig.~\ref{Fig:0437} for further details.}
 \label{Fig:1857}
\end{figure}

Our results, shown in Fig.~\ref{Fig:1857} are in excellent agreement
with those of \citet{ovhb04} showing a relatively weakly polarized
main pulse and interpulse, both of which have multiple
components. \citet{ts90} observed a weak steep-spectrum component just
preceding the main pulse and we see this component as well (at phase
$-0.1$ in Fig.~\ref{Fig:1857}). The PA variation is complex across the
main pulse and completely inconsistent with the RVM. There is evidence
for an orthogonal mode transition near the highest peak of the
interpulse \citep[cf.,][]{xkj+98}.
 
\subsection{PSR J1909$-$3744}

\begin{figure}
\centerline
{
 \psfig{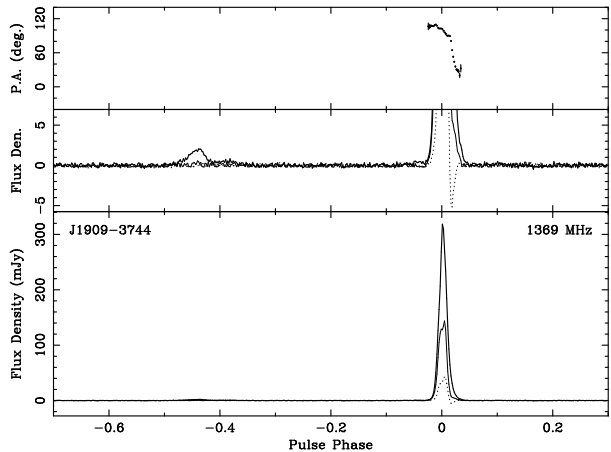}
}
\caption{Polarization profiles for PSR J1909$-$3744 at 1369 MHz. See
  Fig.~\ref{Fig:0437} for further details.}
 \label{Fig:1909}
\end{figure}

The mean pulse profile for this pulsar has a narrow main pulse and a
weak feature preceding the main pulse by about 0.45 in phase. Our
observations, shown in Fig.~\ref{Fig:1909}, are generally consistent
with those of \citet{ovhb04} but show some additional features. The
``interpulse'' feature is clearly seen to consist of two
components and there is evidence for a weak shelf of emission
immediately preceding the main pulse. The main pulse is moderately
polarized and has a PA variation which is almost consistent with the
RVM and a very small impact parameter. However, the total PA swing is
closer to $90\degr$ rather than the expected $180\degr$. This and the
coincident sense reversal of circular polarization suggest that the
rapid swing is actually a smeared orthogonal transition. The smearing
is not instrumental and must be intrinsic to the emitted radiation.

\subsection{PSR J1939+2134 (PSR B1937+21)}

\begin{figure}
\centerline
{
 \psfig{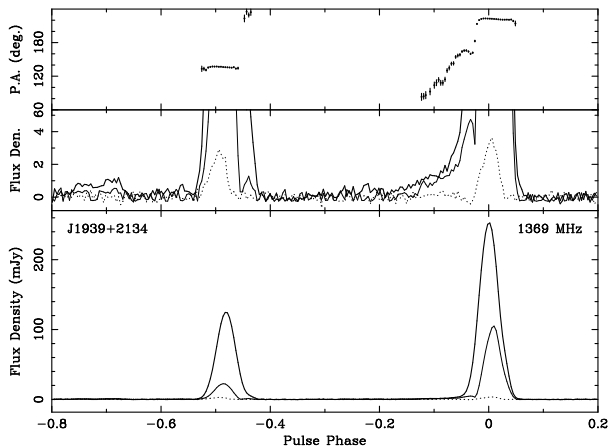}
}
\caption{Polarization profiles for PSR J1939+2134 at 1369 MHz. See
  Fig.~\ref{Fig:0437} for further details.}
 \label{Fig:1939}
\end{figure}

The mean pulse profile of PSR J1939+2134, the first-discovered
MSP \citep{bkh+82}, has a main pulse and a relatively strong
interpulse separated by approximately $\sim 172\degr$ or 0.48 in pulse
phase. Because of the high DM/P of this pulsar, our observations shown
in Fig.~\ref{Fig:1939} are significantly affected by DM smearing (see
Table~\ref{tb:obs}). As a result, we do not see the secondary maxima at
the trailing edges of both pulses seen by \citet{ts90}, \citet{stc99}
and \citet{ovhb04} all of whom used coherent dedispersion. However,
the high sensitivity of our observations reveals previously undetected
pulsed emission preceding both the main pulse and the interpulse by
about 0.2 in phase. This emission has a peak flux density about 0.5\%
of that of the main pulse and appears to have multiple components. At
such a low level, one is concerned that otherwise undetected radio
frequency interference (RFI) or instrumental problems may be
responsible. However, when the data set is split in both frequency and
time, the low-level components are visible in all of the partial data
sets. Although low-level features have been detected in a number of
pulsars in this study, in general they do not have the same
relationship to the main pulse emission as is seen for PSR
J1939+2134. Furthermore, for PSR J1939+2134, the emission preceding
the interpulse has a different profile to that preceding the main
pulse. For these reasons, we believe that the low-level emission seen
in Fig.~\ref{Fig:1939} is real and not either an instrumental artifact
or due to RFI.

The polarization properties shown in Fig.~\ref{Fig:1939} are generally
consistent with previously published polarimetry
\citep{ts90,stc99,ovhb04} although we see more significant
left-circular emission, especially in the main pulse. All three of the
earlier papers referred to above find evidence for an orthogonal mode
transition near the leading edge of the main pulse. However, our data
suggest that the jump is closer to $60\degr$ and therefore not
orthogonal. On the other hand, there is some evidence for a reversal
in the sense of circular polarization at the phase of the PA
transition. We also see a previously undetected orthogonal transition
near the trailing edge of the interpulse.  \citet{ts90} and
\citet{stc99} find evidence for an orthogonal jump at low frequencies
near the leading edge of the interpulse, but no evidence for jumps at
frequencies around 1400 MHz.

\subsection{PSR J2124$-$3358}

\begin{figure}
\centerline
{
 \psfig{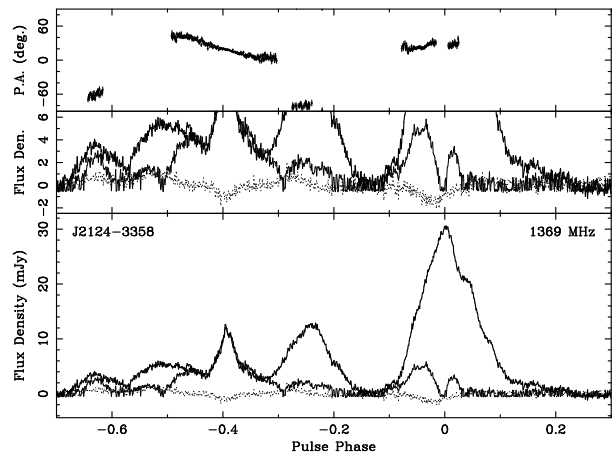}
}
\caption{Polarization profiles for PSR J2124$-$3358 at 1369 MHz. See
  Fig.~\ref{Fig:0437} for further details.}
 \label{Fig:2124}
\end{figure}

As Fig.~\ref{Fig:2124} shows, this pulsar has a extremely complicated
multi-component mean pulse profile with emission covering nearly the
entire pulse period.  The polarization profiles presented here are in
good agreement with those presented by \citet{mh04} and \citet{ovhb04}
for frequencies around 1400 MHz.  The fractional linear polarization
changes greatly across the profile with the emission near phase $-0.4$
being essentially 100\% linearly polarized, whereas, on average, the
profile is about 18\% polarized (Table~\ref{tb:poln}).  As discussed
by \citet{mh04}, the fact that no ``super-polarization'' is seen
suggests that there is no strong steady emission from the pulsar. We
confirm that weak but significant right-circular polarization is
present for the components near phase zero and phase $-0.4$ in
Fig.~\ref{Fig:2124}. In agreement with previous work, an orthogonal
mode transition is seen near phase $-0.3$. If the PAs for emission
between phases of $-0.3$ and $-0.2$ and those for the leading portion
of the profile \citep[to phase $-0.5$, see][]{mh04} are raised by
$90\degr$, then the PA variation approximates an RVM variation for an
outer line of sight with the magnetic axis aligned at about $25\degr$
to the rotation axis. Such a nearly aligned system may explain the
fact that we see emission over most if not all of the pulse period.

\subsection{PSR J2129$-$5721}

\begin{figure}
\centerline
{
 \psfig{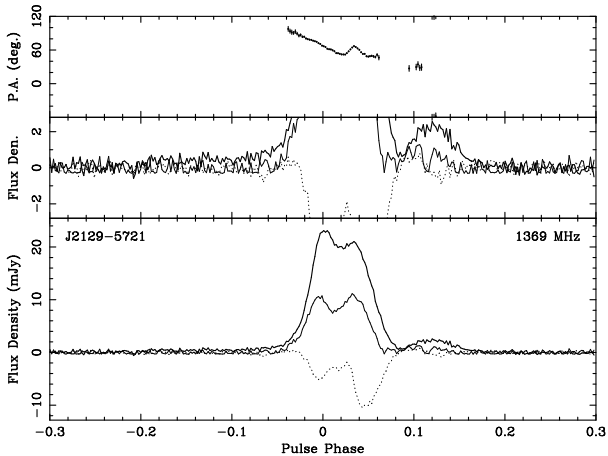}
}
\caption{Polarization profiles for PSR J2129$-$5721 at 1369 MHz. See
  Fig.~\ref{Fig:0437} for further details.}
 \label{Fig:2129}
\end{figure}

This pulsar has been studied at 659 MHz and 1331 MHz by \citet{mh04}
and at 1373 MHz by \citet{ovhb04}. The results shown in
Fig.~\ref{Fig:2129} are in agreement with these earlier results, but
show more detail. All observations detect the postcursor and a weaker
leading shelf of emission, but the present results show that this
extends to at least phase $-0.2$, giving the pulse an overall width of
at least 0.4 in phase. The PA mostly decreases through the pulse, but
the present results clearly show a PA ``glitch'' at the phase of the
trailing main-pulse component. The degree of circular polarization is
unusually high, especially for the trailing main-pulse component. On
average it is 23\% right-circularly polarized, the second highest in
Table~\ref{tb:poln} after PSR J1603$-$7202. 

\subsection{PSR J2145$-$0750}

\begin{figure}
\centerline
{
 \psfig{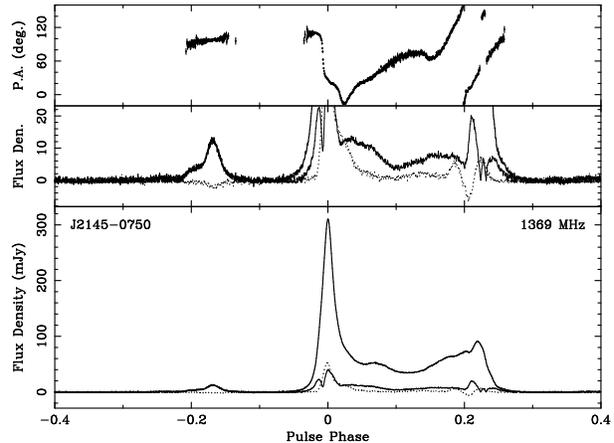}
}
\caption{Polarization profiles for PSR J2145$-$0750 at 1369 MHz. See
  Fig.~\ref{Fig:0437} for further details.}
 \label{Fig:2145}
\end{figure}

The polarimetry of this relatively strong MSP (mean flux density at
9.3 mJy, third to PSRs J0437$-$4715 and J1939+2134, Table~\ref{tb:poln}) has
been studied by a number of groups \citep{stc99,ovhb04,mh04}. Our
results, presented in Fig.~\ref{Fig:2145}, are generally in good
agreement with these earlier authors, but show more detail. Both
\citet{mh04} and \citet{ovhb04} found a weak bridge of emission
between the leading and main pulse components, but we do not see a
bridge. The apparent bridge in the earlier results is probably an
artifact arising from the 2-bit digitization of the input bandpass
employed in these systems. Our results show structure in the leading
component with at least two sub-components. In agreement with earlier
work, this component is essentially 100\% linearly polarized and has
almost constant PA. The PA variation in the rest of the profile is
very complex with regions of increasing and decreasing PA. There are
remarkably close orthogonal transitions (up and back) at phases
0.223 and 0.231 with corresponding dips in $L$. There are also sense
changes in $V$ at phases 0.197 and 0.215, but these do not seem to be associated
with the orthogonal mode jumps which are slightly later in phase.

\section{Rotation measures}\label{sec:rm}
Many of the PPTA pulsars either did not have published RMs or the
published values were insufficiently accurate for summing our data across
frequency to form final profiles. As described in
Section~\ref{sec:obs}, the variable ionospheric RM was subtracted from
the measured values to give an accurate estimate of the interstellar
RM for each pulsar. Table~\ref{tb:RMs} gives our estimates of the
interstellar RM together with the Galactic coordinates and distance
(estimated from the DM using the \citet{cl02} model) of the pulsar and
previously published RM values. In some cases the previously published
values do not have the ionospheric component removed but this would
generally be smaller than the quoted uncertainty. We also give the
mean value of the line-of-sight component of the interstellar magnetic
field in microgauss weighted by the local electron density along the
path (Equation~\ref{eq:bpar}). Negative values correspond
to fields directed away from Earth.

\begin{table*}
\begin{center}
\centering
\caption{Interstellar rotation measures for 20 millisecond pulsars}
\label{tb:RMs}
\begin{threeparttable}
\begin{tabular}{ccccccc}
\hline
{PSR} & Gal. $l$ & Gal. $b$ & Dist. & RM (This work)  &RM (Prev. Publ.) & $\langle B_{||}\rangle$  \\
 & (deg) & (deg) & (kpc) & (rad m\textsuperscript{-2})  &(rad m\textsuperscript{-2}) & ($\mu$G) \\
\hline
J0437$-$4715 & 253.39 & $-$41.96 & 0.16 &  0.0 $\pm$ 0.4  & 1.5 $\pm$ 5\tnote{a}   &$-$0.0 $\pm$ 0.19  \\
J0613$-$0200 & 210.41 & $-$9.30  & 1.25 &  9.7 $\pm$ 1.1    & 19 $\pm$ 14\tnote{b}   &   0.31 $\pm$ 0.03  \\
J0711$-$6830 & 279.53 & $-$23.28 & 0.86 &  21.6 $\pm$ 3.1    & 67 $\pm$ 23\tnote{b}   &   1.45 $\pm$ 0.21  \\
J1022+1001   & 231.79 & $+$51.10 & 0.56 &  $-$0.6 $\pm$ 0.5  & ---                    &$-$0.07 $\pm$ 0.06  \\
J1024$-$0719 & 251.70 & $+$40.52 & 0.53 &  $-$8.2 $\pm$ 0.8  & ---                    &$-$1.56 $\pm$ 0.15  \\ \\
J1045$-$4509 & 280.85 & $+$12.25 & 0.30 &  92.0 $\pm$ 1.0    & 82 $\pm$ 18\tnote{b}   &   1.95 $\pm$ 0.02  \\
J1600$-$3053 & 344.09 & $+$16.45 & 5.00 &  $-$15.5 $\pm$ 1.0 & ---                    &$-$0.36 $\pm$ 0.02  \\
J1603$-$7202 & 316.63 & $-$14.50 & 1.17 &  27.7 $\pm$ 0.8    & 20.1 $\pm$ 5\tnote{b}  &   0.90 $\pm$ 0.03  \\
J1643$-$1224 & 5.67   & $+$21.22 & 0.45 & $-$308.1 $\pm$ 1.0 &$-$263 $\pm$ 15\tnote{b}&$-$6.08 $\pm$ 0.02  \\
J1713+0747   & 28.75  & $+$25.22 & 1.05 &  8.4 $\pm$ 0.6     & ---                    &   0.65 $\pm$ 0.05  \\ \\
J1730$-$2304 & 3.14   & $+$6.02  & 0.53 & $-$7.2 $\pm$ 2.2   & ---                    &$-$0.92 $\pm$ 0.28  \\
J1732$-$5049 & 340.03 & $-$9.45  & 1.41 &  $-$8.5 $\pm$ 6.7 & ---                    &$-$0.18 $\pm$ 0.15  \\
J1744$-$1134 & 14.79  & $+$9.18  & 0.42 &  $-$1.6 $\pm$ 0.7  & ---                    &$-$0.63 $\pm$ 0.27  \\
J1824$-$2452 & 7.80   & $-$5.58  & 4.90 & 77.8 $\pm$ 0.6     & 1 $\pm$ 12\tnote{c}    &   0.80 $\pm$ 0.01  \\
J1857+0943   & 42.29  & $+$3.06  & 0.91 &  16.4$\pm$ 3.5    & 53 $\pm$ 9\tnote{d}    &   1.52 $\pm$ 0.32  \\ \\
J1909$-$3744 & 359.73 & $-$19.60 & 1.27 & $-$6.6 $\pm$ 0.8   & ---                    &$-$0.78 $\pm$ 0.09  \\
J1939+2134   & 57.51  & $-$0.29  & 8.33 &  6.7 $\pm$ 0.6     & $-$10 $\pm$ 9\tnote{c} &   0.12 $\pm$ 0.01  \\
J2124$-$3358 & 10.93  & $-$45.44 & 0.32 &  $-$5.0 $\pm$ 0.9  & 1.2 $\pm$ 1\tnote{b}   &$-$1.34 $\pm$ 0.24  \\
J2129$-$5721 & 338.01 & $-$43.57 & 0.53 &  23.5 $\pm$ 0.8    & 37.3 $\pm$ 2\tnote{b}  &   0.91 $\pm$ 0.03  \\
J2145$-$0750 & 47.78  & $-$42.08 & 0.62 &  $-$1.3 $\pm$ 0.7  & 12 $\pm$ 8\tnote{b}    &$-$0.18 $\pm$ 0.1  \\
\hline
\end{tabular}
   \begin{tablenotes}
     \item References: (a) \citet{nms+97}; (b) \citet{mh04}; (c) \citet{rl94}; (d) \citet{hml+06}.
    \end{tablenotes}
\end{threeparttable}
\end{center}
\end{table*}

In most cases our RM measurements are close to previous measurements
within the uncertainties. However, PSR J1824$-$2452 (PSR B1821-24)
stands out; in this case, it seems likely that the previously
published result was incorrect. For several other pulsars, the results
differ by more than the combined error. There are several possible
reasons for this. Perhaps the most likely is that unknown systematic
errors, most probably in calibration, resulted in underestimation of
the RM uncertainties. Alternatively, it is possible that in some cases
there has been a real change in RM since the earlier
measurements. Continuing PPTA observations will help to establish if
such variations exist.

\section{DISCUSSION AND CONCLUSIONS} \label{sec:discn} 

We have presented new and improved polarization profiles for the 20
MSPs observed as part of the Parkes Pulsar Timing Array
project. Because of the frequent and relatively long observations
needed by the PPTA project, our polarization profiles generally have
very high S/N ratios compared to earlier results. Also, we have
employed improved signal processing procedures, including corrections
for feed cross-coupling and ionospheric Faraday rotation. As a result,
we have not only defined the polarization properties more accurately,
but also revealed previously unknown profile features in many of the
pulsars. 

In a number of cases, the newly detected profile features greatly
extend the range of pulse phase over which emission is detected. For
example, in PSR J1045$-$4509, the detection of a low-level leading
component and bridge emission joining it to the main pulse has
increased the overall observed width of the pulsed emission by nearly
a factor of three, from about 0.25 to about 0.7 in phase. PSR
J1939+2134 (PSR B1937+21) is an interesting case in which the newly
detected emission changes the pulse profile from having just a main
pulse and interpulse separated by close to 0.5 in phase and both
relatively narrow, to a wide profile covering about 80\% of the pulse
period. In some ways, the situation is similar to that of the Crab
pulsar, where pulse components with high linear polarization are
detected at phases very different to those of the main pulse and
interpulse \citep{mh96,mh99}. These pulse components are only seen at
high radio frequencies, but the Crab pulsar also has a highly
polarized precursor pulse which is only seen at low radio frequencies
\citep{man71c,mh96}. Some, but not all, of the outlying components
detected in this study are highly linearly polarized but those in PSR
J1939+2134 are not.

\begin{figure}
\centerline
{
 \psfig{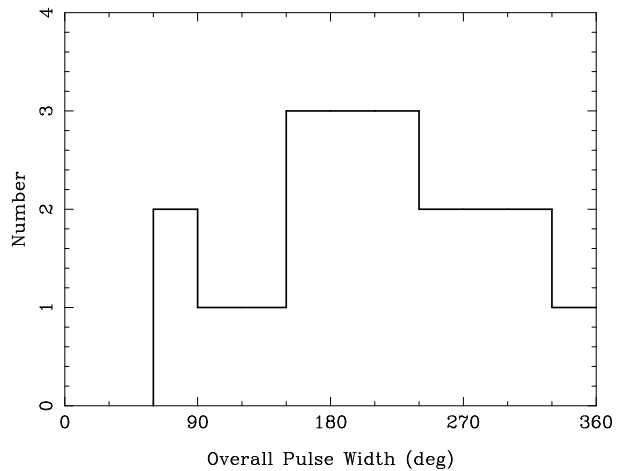}
}
\caption{Histogram of overall pulse widths for the 20 PPTA pulsars}
 \label{fg:widths}
\end{figure}

Table~\ref{tb:poln} gives the overall pulse width, that is, the total
longitude range over which significant pulsed emission is seen,
for each of the PPTA pulsars. The distribution of these widths is
plotted in Fig.~\ref{fg:widths}. Only seven of the 20 pulsars have
emission spanning less than half of the period and five have emission
over more than three-quarters of the period. Only in a few of the
pulsars does the pulse profile have the form of a main pulse and
interpulse separated by close to half the period. Even when there are
strong pulse components that can be labelled in this way, there is
generally emission at other pulse phases as well.

Most normal (non-millisecond) pulsars have mean pulse profiles that can be
described within the core--cone framework with one to five clearly
identifiable pulse components or peaks within a relatively narrow
longitude range, typically about $20\degr$. Although the relative
amplitude of the various components varies widely, leading to the
notion of partial cones \citep{lm88}, there often appears to be a
symmetry about a central phase, leading to the idea of nested cones
\citep[e.g.,][]{ran90,gg03}. However, observations with high S/N ratio
often show a more complex profile structure with multiple components
\citep[e.g., PSR B0740$-$28 in][]{kra94} which are not easily
accommodated in the nested-cone model. This and the wide variation of
component amplitudes led to the idea of ``patchy'' emission beams
\citep{lm88,hm01}. Pulsars with interpulses and/or wide profiles form
another group of multiple component pulsars. Only a few normal pulsars
have these features, with less than 2\% showing interpulse emission
\citep{wj08a} and most of this 2\% have short periods and high
spin-down luminosities. Many have the main pulse and interpulse
separated by close to $180\degr$, but some have smaller spacings
and/or bridges of emission between the two components. Good examples
are the Crab pulsar as discussed above, PSR B0950+08 and PSR B1929+10
\citep{hf86}. Some interpulse pulsars also have weaker pulse
components, often highly linearly polarized, separated from the main
pulse and interpulse and usually termed ``precursors'' or
``postcursors''. Examples are the Crab pulsar again, PSR B0823+26
\citep{hf86,rr95b} and PSR B1822-09 \citep{mhm80,jkk+07}.

As mentioned above, wide profiles and interpulses are much more common
in MSPs, with more than 60\% of the present sample showing these
features. PSR J1045$-$4509 (Fig.~\ref{Fig:1045}) is a good example of
a nominal interpulse joined to the main pulse by a bridge of
emission. Highly polarized precursors and postcursors are observed in
PSR J1024$-$0719 (Fig.~\ref{Fig:1024}), PSR J1603$-$7202
(Fig.~\ref{Fig:1603}), PSR J1713+0747 (Fig.~\ref{Fig:1713}) and PSR
J1730$-$2304 (Fig.~\ref{Fig:1730}). Multiple components are ubiquitous
in MSPs. Some components are obvious as clearly resolved peaks in the
pulse profile, but others are just points of inflection in the profile
and probably some are not obvious in the profile at
all. Table~\ref{tb:poln} gives an estimate of the minimum number of
identifiable components in the PPTA MSPs based on peaks or points of
inflection in the pulse profiles. 

In most cases, the observed PA variations across these wide profiles are
incredibly complex, even after allowance for orthogonal mode
transitions, and are not even approximately described by the
RVM. Exceptions are PSRs J1024$-$0719, J1600$-$3053 and J1732$-$5049
which have relatively flat PA profiles, and PSRs J1022+1001,
J1744$-$1134 and J2124$-$3358 which have reasonably smooth PA
variations that are consistent with the RVM. Except for PSR
J1022+1001, RVM fits to these PA variations imply large impact
parameters and/or nearly aligned magnetic and rotational
axes. For PSR J2124$-$3358, a nearly aligned model may well be the
correct interpretation since it has emission across essentially the
whole period, but in other cases it is not so clear. 

For most of the pulsars observed, the PA variations are complex with
disconnected PA segments, often with opposite slopes in neighbouring
regions (e.g., PSR J0613$-$0200, Fig.~\ref{Fig:0613}; PSR
J1603$-$7202, Fig.~\ref{Fig:1603}; PSR J1730$-$2304,
Fig.~\ref{Fig:1730}; PSR J2145$-$0750, Fig.~\ref{Fig:2145}) and
non-orthogonal PA jumps (e.g., PSR J0437$-$4715, Fig.~\ref{Fig:0437};
PSR J0711$-$6830, Fig.~\ref{Fig:0711}; PSR 1713+0747,
Fig.~\ref{Fig:1713}). Within a single component, the PA variation is
usually continuous and monotonic (orthogonal jumps excepted). These
results suggest that overlapping components may originate in different
locations within the magnetosphere. Supporting evidence for
quasi-independent emission regions comes from phase-resolved spectral
index plots which show different spectral indices for different
components \citep[e.g.,][]{mh04}. \citet{drd10} have also suggested
the idea of independent emission from different parts of the pulsar
magnetosphere, in their case, from thin plasma streams emitting
coherent curvature radiation. In this model, the symmetry axis for PA
variations is not the magnetic axis (as in the dipole polar-cap model)
but the plasma-stream direction. This could certainly account for the
non-RVM PA variations we observe. However, we do not find evidence for
the common occurence of the double features they identify with the basic
curvature radiation beam.

The very wide MSP profiles we observe (Fig.~\ref{fg:widths}) imply
large emission heights in the simple polar-cap model. In fact, for a
pulsar period of 4~ms and an overall pulse width of $200\degr$ (about
the median values for our sample), Equation~\ref{eq:beam} implies
$r/R_{LC} \gapp 1.3$. One could invoke small inclination angles, but
it is more likely that the equation is simply not
applicable. Similarly, the standard methods for estimating emission
heights cannot be applied to MSPs. They do not exhibit
radius-frequency mapping of component separations and hence that
method \citep{cor78} cannot be used. Likwise, the method of
\citet{bcw91}, which relies on identifying a relativistic
aberration/retardation shift between the pulse profile and the PA
variations, cannot be applied because of the general lack of symmetry
about a central longitude for one or both properties. For example,
PSRs J0437$-$4715 and J2145$-$0750 both have reasonably symmetric
pulse profiles, but in both cases it is impossible to determine any
symmetry centre for the PA variations. Relativistic methods which rely
on a height difference between core and conal components
\citep{gg03,drh04} also cannot easily be applied to MSPs since it is
generally impossible to uniquely identify components as either core or
cone.

Based on associations with gamma-ray detections, \citet{man05} and
\citet{rmh10} argued that radio beams from young and millisecond
pulsars are emitted from the outer magnetosphere and that caustic
effects are important in defining the observed pulse profile. By
analogy with gamma-ray pulse profiles \citep[e.g.,][]{dr03,wrwj09}
such effects may account for the broad frequency-independent pulse
profiles observed in MSPs, including interpulses and connecting
bridges. The results presented in this paper support this idea. The
observed pulse profiles are too wide and too complex to be simply
accounted for as emission tangential to the open field lines at low
altitudes in the pulsar magnetosphere. Observed PA variations are also
generally inconsistent with this interpretation. For example, the
extended low-level emission observed for PSR J1939+2134
(Fig.~\ref{Fig:1939}) shows that main pulse and interpulse in this
pulsar are not simple polar beams from opposite magnetic poles as was
originally thought. In this context, it is important to note that, for
both the Crab pulsar and PSR J1939+2134, giant radio pulses occur at
phases closely related to the phase of the high-energy emission and to
the main radio peaks \citep{aaa+10a,mjr10,chk+03} also supporting the
idea of the radio emission originating in the outer magnetosphere. 

\citet{dwd10} discussed the caustic deformation of radio pulse
profiles by aberration and retardation, albeit to a more modest degree
than that proposed above. Trailing sides of profiles are compressed in
width and enhanced in amplitude whereas leading parts are spread out
and reduced in amplitude. They apply these ideas to the main pulse of
the MSP J1012+5307, which has a strong trailing component and a weaker
more extended leading part, and can model the observed profile with a
patchy underlying emission and a small relativistic deformation, with
the emission originating between the stellar surface and twice that
radius. Of the pulsars we observed, only PSR J1022+1001 and maybe PSR
J1600$-$3035 have similar profiles with a relatively strong and narrow
trailing component. It is clear that these ideas are not generally
applicable to MSPs. 

\citet{jw06} suggested that emission from many young pulsars arises
from the outermost open field lines at relatively high altitudes,
between 5 and 15 per cent of the light-cylinder radius. We are
proposing that for MSPs (and some young pulsars as well) the emission
zone extends much higher, up to or even beyond the null-charge
surface, with caustic deformations a dominant influence on the profile
morphology. To account for the multiple components and complex PA
variations observed in MSPs, much structure is required in the
underlying emission. This could originate from isolated plasma streams
as suggested by \citet{drd10} with the observed PA determined by the
projection of the stream axis or by the local magnetic field
direction. In the outer magnetosphere, such currents will
significantly deform the basic dipolar field, leading to the observed
small-scale PA variations and discontinuities. It is worth noting
though that the observed complex profile and PA variations are
extremely stable over timescales of decades (compare the PSR
J0437$-$4715 profiles in \citet{nms+97} and in
Fig.~\ref{Fig:0437}). Observations of nulling and mode-changing in
normal pulsars \citep[e.g.,][]{wmj07,lhk+10} show that magnetospheric
currents can vary on short timescales, but to date there is little
evidence that such phenomena exist in MSPs. Even in normal pulsars,
the properties of a given mode seem quite stable. Inwardly directed
emission beams from charges flowing toward the neutron star
\citep{gjk+94,dfs+05} provides another possible explanation for the
multiple and overlapping components of MSP pulse profiles.

Despite the generally more complex pulse profiles of MSPs compared to
normal pulsars, there are many similarities in the emission
characteristics. Fractional linear polarizations are often very high
for both classes of pulsar, orthogonal-mode transitions are ubiquitous
in both classes and circular polarization is often seen, but is almost
always weaker than the linear polarization. Truly orthogonal
polarization transitions (both circular and linear changing sign) do
seem more common in MSPs with at least five clear examples (in PSRs
J0437$-$4715, J1600$-$3053, two in J1643$-$1224 and J1713+0747) among
the 20 PPTA pulsars presented here. Overall though, the similarities in
polarization behaviour outweigh the differences, strongly suggesting
that the basic radio emission mechanism is the same for both normal
pulsars and MSPs.

We derive RMs for all 20 pulsars, eight of which have no previously
published RMs. For those with previously published RMs, our
measurements have smaller estimated uncertainties and we believe they are more
accurate than the previous values. For three pulsars (J1824$-$2452,
J1939+2134 and J2145$-$0750) our measurements differ significantly,
indicating systematic errors, most probably in the earlier
measurements. RMs and implied mean values of the line-of-sight
component of the interstellar magnetic field are small for
high-latitude and more distant pulsars in accord with earlier results
\citep[e.g.,][]{hml+06}. Most measured values are less than or about
$1\;\mu$G, but that for PSR J1643$-$1224 at $6\;\mu$G is an
exception. The estimated distance for this pulsar is only about 0.5
kpc and the path to it traverses the North Polar Spur
\citep{wol07}. The large mean field strength almost certainly results
from compressed interstellar fields in the shell(s) which form this
prominent feature of the Galactic radio background.

\subsection*{ACKNOWLEDGMENTS}
The Parkes Pulsar Timing Array (PPTA) project is a collaboration
between a number of groups both in Australia in other countries
established with the support of RNM's Australian Research Council
Federation Fellowship (\#FF0348478). GH is the recipient of an
Australian Research Council QEII Fellowship (\#DP0878388). WMY is 
supported by NSFC project 10673021, the Knowledge Innovation Program 
of the Chinese Academy of Sciences, Grant No. KJCX2-YW-T09 and National 
Basic Research Program of China (973 Program 2009CB824800). We thank
members of the PPTA collaboration who assisted with the observations
reported in this paper. We thank Andrew Gray and Ken Tapping for
making the Penticton ionospheric modelling program available to
us. The Parkes radio telescope is part of the Australia Telescope,
which is funded by the Commonwealth of Australia for operation as a
National Facility managed by the Commonwealth Scientific and
Industrial Research Organisation.


\label{lastpage}

\end{document}